\documentstyle[11pt]{article}
\def\thebibliography#1{%\section*{REFERENCES}
   \def\item{\par\hangindent1em\parindent0em}
}

\begin{document}

\centerline{}

\centerline{}

\centerline{}

\centerline{}

\centerline
{\LARGE \bf COMET HAZARD TO THE EARTH} 

\centerline{}

\centerline{}

\centerline
{S. I. Ipatov}

\centerline{}

\centerline
{\it Institute of Applied Mathematics, Miusskaya sq. 4, Moscow 125047, Russia;
ipatov@keldysh.ru}

\centerline{}

\centerline{}

\centerline{\bf ABSTRACT}

\centerline{}

\noindent
Migration of trans-Neptunian objects to the Earth is considered. Due to the gravitational influence of large trans-Neptunian objects and mutual collisions, some objects can get such orbits, from which they can be moved inside the Solar System under the gravitational influence of planets. About 10-20\% or even more 1-km Earth-crossers could have come from the Edgeworth-Kuiper belt and can move in Jupiter-crossing orbits. The number of former trans-Neptunian objects that cross the Earth's orbit moving in orbits with aphelia inside Jupiter's orbit can be of the same order of magnitude. A total mass of icy planetesimals delivered to the Earth during formation of the giant planets can be about the mass of the Earth's oceans.

\noindent {\bf Keywords}: trans-Neptunian objects, near-Earth objects, migration

\medskip
Paper submitted to "Advances in Space Research" (in press).

%\centerline{}

%\centerline{}

%\centerline{}
\newpage

\noindent
{\bf INTRODUCTION}

%\centerline{}

Near-Earth objects (NEOs), i.e., objects with perihelion distance $q \le 1.3$ AU and aphelion distance $Q \ge 0.983$ AU, are considered to have a chance to collide the Earth in some future. There are more than 30 NEOs larger than 5 km in diameter, about 1500 NEOs larger than 1 km, and 135,000 larger than 100 m (Rabinowitz {\it et al.}, 1994). About half of NEOs are Earth-crossers.
It is considered that diameter of the object that caused the Tunguska event in 1908 was about 60-70 m. The main sources of NEOs are considered to be the main asteroid belt and the Edgeworth--Kuiper belt (EKB). Some Earth-crossers are long-period comets, which came from the Oort cloud located at a distance $(2-3)\cdot10^4 - 10^5$ AU from the Sun. It is not yet clear, how many NEOs came from the EKB. Some scientists (Farinella {\it et al.}, 1993) considered that most of meteorites and NEOs are asteroidal fragments, other (Wetherill, 1988) supposed that half of NEOs are former short-period comets, and Eneev (1980) considered that most NEOs came from the EKB. A cometary origin of NEOs, as well as meteorites, was suggested by \"Opik (1963). Fernandez (1980) first supposed that short-period comets came from the trans-Neptunian belt. Icy bodies can also migrate inside the Solar System from the regions located between the EKB and the Oort cloud. According to Weissman (1995), the dynamically inactive region beyond 45 AU may extend out to 1000 AU or even more and contain up to several times $10^{13}$ objects with a total mass of several hundred Earth masses.

Some bodies leave the main asteroid belt via different mean motion and 
secular resonances (Gladman {\it et al.}, 1997). Objects that came from 
the main asteroid belt got in the resonant regions, from which they were 
delivered to a near-Earth space, mainly via collisions with other asteroids 
(for small asteroids also due to the Yarkovsky effect). 
%and usually doesn't exceed 5--10 km in diameter. 
Ipatov (1995b) obtained that only a few percents of asteroids could
get into the main resonances with Jupiter (3/1, 5/2, 7/3, and 2/1)
during the last 4 Gyr due to the mutual gravitational
influence of asteroids. Migliorini {\it et al.} (1998), Morbidelli
and Nesvorny (1999) showed that mean motion resonances with Mars (3/5, 7/12, 4/7, 5/9, 7/13, and 1/2), mean motion resonances with Jupiter (7/2 and 10/3),
and three-body mean motion resonances between Jupiter, Saturn, and
the asteroid, or Mars, Jupiter, and the asteroid could deliver
material to Mars and Earth. As the number of resonances delivering
bodies to the Earth is not small, even small variations in semimajor
axes can put some asteroids in the resonances. So the role of mutual
gravitational influence in transportation of asteroids to the Earth
may not be very small. According to Wetherill (1988) and Weissman 
{\it et al.} (1989), it is difficult to explain the number of NEOs
and features of their orbits (for example, their mean inclinations, 
which are larger than those in the main asteroid belt), if one considers only asteroidal sources. Wetherill (1991) considered that NEOs should come from the trans-Neptunian belt, but not from the Oort cloud in order to supply present inclinations of orbits of NEOs. To our opinion, the mean inclination of those NEOs that came from the main asteroid belt can be the same as that of all NEOs ($15^\circ$) and can be larger than the mean inclination of main-belt asteroids ($10^\circ$), because these objects could increase their inclinations when they were in resonances.

In April 2001, 369 typical trans-Neptunian objects (TNOs) and 67 Centaurs/SDOs were discovered. Typical TNOs are often called EKB objects (EKBOs). EKBOs have semimajor axes between 35 and 50 AU, and their eccentricities are less than 0.4. As there is no exact boundary between Centaurs
%, which cross orbits of the giant planets, but are not Jupiter-crossers, 
and SDOs (scattered disk objects), these objects are listed in one table. Most of SDOs have $a>50$ AU. The largest TNO (2000 WR106, i.e., Varuna (20000), $a=43.3$ AU, $e=0.056$, $i=17^\circ$) has diameter about 1000 km. Most of other TNOs has diameters $d\sim 100-400$ km. About 70,000 of such sized bodies were inferred (Jewitt {\it et al.} 1996, Levison and Duncan 1997). The total mass of EKBOs $M_{EKB}$ is estimated to be about $0.06m_\oplus$ to $0.25m_\oplus$, where $m_\oplus$ is the mass of the Earth (Jewitt {\it et al.}, 1996). The total number of bodies having diameter $d\ge1$ km within the $30\le a \le50$ AU is estimated to be about $10^{10}$ (Jewitt {\it et al.}, 1996), $10^{11}$ (Jewitt, 1999), or $5\cdot 10^9$ (Jewitt and Fernandez, 2001). The mean values of eccentricity and inclination for first 181 TNOs (with $a\le50$ AU) are equal to $0.09$ and $8.1^\circ$, respectively. For 34 SDOs with $a>50$ AU and $q \ge 30$ AU, these values were $0.5$ and $15^\circ$. A total mass of SDOs that move in eccentric orbits between 40 and 200 AU is estimated to be equal to $0.5M_\oplus$ (Luu {\it et al.}, 1997) or to $0.05M_\oplus$ (Trujillo {\it et al.}, 2000). In the latter paper the number of ``scattered'' objects with $d > 100$ km is considered to be about $3 \cdot 10^4$. 
%Bodies migrated from the EKB can consist mainly of water and volatiles.

Long-period and sungrazing comets also deliver material to the
terrestrial planets. About 80\% of all known comets are long-period comets, more than 80 members of the family being sungrazing comets. More than half of the close (up to 0.102 AU) encounters of comets with the Earth belong to long-period comets (http://cfa-www.harvard.edu/iau/lists/ClosestComets.htm). Thus, the number of collisions of active long-period comets with the terrestrial planets may be of the same order of magnitude as that for active short-period comets, but short-period comets supply a larger number of extinct comets. Some long-period comets are as large as the Hale--Bopp comet, which had diameter of about 30 km, perihelion distance $q = 0.914$ AU and moved almost perpendicular to the plane of the Earth's orbit. The mean impact probability of a long-period comet with the Earth per revolution is estimated as $(2-3)\cdot10^{-9}$ (Marsden and Steel, 1994).

Active and extinct periodic comets may account altogether for about 20\% of the production of terrestrial impact craters larger than 20 km in diameter (Shoemaker {\it et al.}, 1994). About 40 active and 800 extinct Earth-crossing Jupiter-family comets with period $P<20$ yr and nuclei $\ge 1$ km, and about 140--270 active Earth-crossing Halley-family comets ($20<P<200$ yr) were estimated by Shoemaker {\it et al.}, some of the latter still assumed to come from the Oort cloud. The study of dynamical evolution of Halley-type comet orbits for about $\pm 1$ Myr showed that some of them got into Earth-crossing orbits for 1/3 to 2/3 of the considered time interval (Bailey and Emel'yanenko, 1996). The probabilities of collisions of short-period comets with Earth and Venus were estimated to be quite similar, whereas that with Mars is an order of magnitude smaller (Nakamura and Kurahashi, 1998). 
%This means that the total mass of the matter delivered from the giant planets region, normalized to the mass of the planet, was nearly the same. 
About 1\% of observed Apollos cross Jupiter's orbit (and additional 1\% of Apollos have aphelia between 4.7-4.8 AU), but these Jupiter-crossers move far from the Earth during most of time, so their actual portion among ECOs is greater than that for observed objects. 
The portion of Earth-crossers among observed Jupiter-family comets is about 10\%.

Beside comets and asteroids, numerous meteoroid--like bodies impact the terrestrial planets. About 98--99\% of such bodies with masses less than 100 g in the vicinity of the Earth are assumed to be of cometary origin, and, in particular, the orbit of the Geminid meteor stream nearly coincides with the orbit of asteroid 3200 Phaethon (Whipple, 1983). Obviously, spectral variations in the collection of meteorites are considerably wider than those of asteroids (Britt {\it et al.}, 1992). Note that water was found in the Monahans (1998) H5 chondrite (Zolensky {\it et al.}, 1999) and it could exist in some other meteorites (Vilas and Zolensky, 1999). Morbidelli {\it et al.} (2000) argues that during the formation of the main asteroid belt the bodies that migrated from this belt could deliver to the Earth even more water than comets.

From the investigation of lunar craters, Hartmann (1995) considers for a relatively uniform size distribution of interplanetary impactors of mixed origins back to 4 Gyr ago. The frequency of  Earth impact by NEOs is estimated to be larger by a factor 2, 14, 24 and 30 as compared to Venus, Mars, Moon and Mercury, respectively (Bottke {\it et al.}, 1994).
Shoemaker {\it et al.} (1990) showed that asteroidal impacts probably dominated the production of craters on the Earth with diameter $D_{cr} < 30$ km, whereas cometary impacts were responsible for the craters with $D_{cr} > 50$ km. This feature can be explained by the idea that TNOs can leave their residence EKB region essentially without collisions, in contrast to main belt asteroid bodies that experienced multiple collisions (Ipatov, 1998, 1999).

In this paper we consider the migration of TNOs and icy planetesimals to the Earth.

\centerline{}

\noindent
{\bf EARLY MIGRATION OF ICY BODIES}
\nolinebreak

%\centerline{}

\noindent
%\underline
{\bf Formation of the Giant Planets and Trans-Neptunian Objects}

%\centerline{}

Estimates of the number of comets migrated to the Earth during the formation of the giant planets depend on the model of their formation. Such bombardment culminated from 4.5 to 4 Gyr ago. It is generally agreed that the total mass of planetesimals in the feeding zone of the giant planets exceeded by a factor of several the total mass of solids that entered into the giant planets (Safronov, 1972). The total mass $M_{UN}$ of planetesimals in the feeding zone of Uranus and Neptune could exceed $100m_\oplus$ (Ipatov, 1987, 1993). Most of these planetesimals could still move in this zone when Jupiter and Saturn had accreted the bulk of their masses. Zharkov and Kozenko (1990) suggested that Uranus and Neptune accreted a considerable portion of their masses in the feeding zone of Saturn, where they aquired also hydrogen envelopes. In a series of numerical experiments, Ipatov (1991, 1993) showed that the embryos of Uranus and Neptune could increase their semimajor axes from $\le 10$ AU to their present values, moving permanently in orbits with small eccentricities, due to the gravitational interactions with the planetesimals that migrated from beyond 10 AU to Jupiter. Later on, the idea of a narrow zone, where all giant planets have been formed, was supported and further developed by Thommes {\it et al.}~(1999).
In my opinion, several giant planets moving in close orbits could be formed 
around another star, but some of these planets, due to their mutual gravitational interactions, could be ejected into hyperbolic orbits and the remaining extrasolar planets (planet) could get large eccentricities.
Ipatov (1987, 1993) showed that during the accumulation of the giant planets the planetesimals with the total mass of several tens $m_\oplus$ entered the trans-Neptunian region. These planetesimals could increase the eccentricities of 'local' TNOs and swept most of these TNOs, which total initial mass could exceed $10m_\oplus$. A small part of such planetesimals could left beyond Neptune's orbit in highly eccentric orbits such as those of SDOs.

Stern (1995, 1996a,b), Stern and Colwell (1997), Davis and Farinella (1997), Kenyon and Luu (1998, 1999) investigated the formation and collisional evolution of the EKB. In their models, the process of accumulation of TNOs took place at small (usually about 0.001) eccentricities and a massive belt. To our opinion, due to the gravitational influence of forming giant planets (our runs showed that maximal eccentricities of TNOs always exceeded 0.05 during 20 Myr under the
gravitational influence of the giant planets) and the mutual gravitational influence of planetesimals, such eccentricities may not exist during all the time needed for the accumulation of TNOs, though gas drag could decrease eccentricities and the gravitational influence of forming planets could be less. 
Eneev (1980) supposed that large TNOs were formed from rarefied dust--gas condensations. We consider that TNOs with $d \ge 100$ km moving now in not very eccentric orbits could be formed directly by the compression of large local rarefied condensations, but not by the accretion of smaller solid planetesimals. Probably, some largest asteroids and planetesimals with $d\ge100$ km in the zone of the giant planets could be formed in the same way. Some smaller objects can be debris of larger objects and other small objects could be formed directly by compression of condensations. Even if at some instant of time at approximately the same distance from the Sun, the dimensions of initial condensations, which had been formed from the dust layer due to gravitational instability, had been almost identical, there was a distribution in masses of the final condensations, which compressed into the planetesimals. As in the case of accumulation of planetesimals, there could be a "run-away" accretion of condensations. 

\centerline{}

\noindent
%\underline
{\bf Early Impactor Delivery}

%\centerline{}

The total mass of the planetesimals from the feeding zone of Uranus and Neptune that collided with Earth from orbits crossing both the orbits of Jupiter and Earth can be estimated by the following formula: $m_{Ec} = M_{UN} p_{J} p_{JE} \Delta t_{E} / T_{E}$, where $p_{J}$ is the fraction of objects from the feeding zone of Uranus and Neptune that reached Jupiter's orbit; $p_{JE}$ is the fraction of Jupiter-crossing objects that reached the orbit of Earth during their lifetimes; $\Delta t_{E}$ is the mean time during which a Jupiter--crossing object crosses the orbit of Earth; and $T_{E}$ is the characteristic time for such an object to impact the Earth. Let us assess quantitatively the factors of the above formula. An estimate of Jupiter-crossing objects can be made on the basis of the orbital evolution of comet P/1996 R2 (Lagerkvist) having the parameters ($a\approx3.79$ AU, $e\approx0.31$, $i\approx2.6^\circ$). This results in $p_{JE}\approx0.2$ and $\Delta t_{E} \approx 5\cdot10^3$ yr (Ipatov and Hahn, 1999). Note that the mean inclination of Earth-crossing objects is about $15^\circ$ and that the larger $a$, the larger $T_E$. At $e=0.7$, $a=3.06$ AU, and $i$ varing from 0 to 30$^\circ$, we have $T_E=400$ Myr. Now, taking $M_{UN}=100m_\oplus$, $p_{JE}=0.2$, $p_{J}=0.5$, $\Delta t_{E}=5\cdot10^3$ yr, and $T_{E}=400$ Myr, we obtain $m_{Ec}=1.2\cdot10^{-4}m_\oplus$. For $\Delta t_E=10^4$ yr, the value of $m_{Ec}$ is greater by a factor 2, both figures being comparable with the mass of water in the Earth's oceans (about $2\cdot10^{-4} m_\oplus$). 
The above quite rough estimates were obtained for the planetesimals which came 
from the feeding zone of Uranus and Neptune and then cross both the orbits 
of Jupiter and Earth. The same order of $m_{Ec}$ can be obtained 
for the planetesimals migrated from the feeding zones of Jupiter and Saturn. 
Probably, even more former planetesimals from the feeding zone of the giant planets could impact the Earth from Encke-type orbits with aphelia $Q\le4.5$ AU. Some objects initially formed beyond 30 AU could also migrate to the Earth during planet formation.

\centerline{}

\noindent {\bf
PRESENT MIGRATION OF COMETS FROM THE EDGEWORTH--KUIPER BELT TO THE EARTH}

%\centerline{}

\noindent
%\underline
{\bf Objects Leaving the Edgeworth--Kuiper Belt}

%\centerline{}

Under the gravitational influence of the giant planets, some TNOs can increase their eccentricities, begin to cross Neptune's orbit, and leave the EKB (Torbett and Smoluchowski, 1990; Gladman and Duncan, 1990; Holman and Wisdom, 1993; Levison and Duncan, 1993; Duncan et al., 1995; Duncan and Levison, 1997; Levison and Duncan, 1997). As for the main asteroid belt (MAB), some TNOs can get orbits, which become planet-crossing after some time, due to collisions. Let us compare the rate of collisions in the MAB and EKB. There are about $10^6$  main-belt asteroids with diameter $d\ge1$ km; the number of asteroids with $d \ge d_* \ge 1$ km is proportional to $d_*^{-\alpha}$, with $\alpha$ between 2 and 2.5 (Binzel {\it et al.}, 1991; Hughes and Harris, 1994). The mass ratio of an asteroid and an impactor for which catastrophic destruction occurs is about $s_0=10^4$ (Petit, 1993; Williams and Wetherill, 1994). In other words, an asteroid of diameter $d$ can be destroyed by an impactor of diameter $d'\ge0.046d$. The characteristic time $T_{cN}$ to destroy an asteroid of diameter $d\ge1$ km through a collision with a body of diameter $d'\ge100$ m ($s_0=10^3$) equals 5 Gyr, and for $d=1$ km and $d'=46$ m ($s_0=10^4$), it is $T_{cN}\approx 1$ Gyr (Ipatov, 1995b). These estimates are of the same order as those by Bottke and Greenberg (2001), who found that a 30-km asteroid is disrupted by some $\approx 2.5$ km projectile in 4.6 Gyr. Canavan (1993) obtained that $\alpha=3$ for NEOs with $d<40$ m and $\alpha=2$ at 40 m $< d<2$ km. If we assume that values of $\alpha$ for the MAB are the same as those for NEOs, then at $s_0=10^4$ for $d=100$ m ($d'=4.6$ m), $d=10$ m ($d'=46$ cm), and $d=1$ m ($d'=4.6$ cm), the values of $T_{cN}$ will be smaller by a factor of $40/4.64\approx 8.6$; 86; and 860, respectively, than those for $\alpha=2$. In this case, the collisional lifetime of a 1-m body is only about 1 Myr; this means that numerous meteoroids are debris of larger asteroid-like bodies. Some small cometary objects and NEOs can be also crashed during their travel via the MAB.

For the EKB (at $M_{EKB} \sim 0.1m_\oplus$), the characteristic time $T_{cN}$ elapsed up to a 
collision of a body with diameter $d \ge D$ with some other 
body $d \ge D_2$ is greater by an order of magnitude than the 
values of $T_{cN}$ obtained for the MAB at the same values of $\alpha$ and $s_0=(D/D_2)^3$, but it is of the same order of magnitude for $D=1$ km if we consider
$10^{10}$ 1-km and $10^{12}$ 100-m TNOs. 
The energy of a collision is proportional to $v_c^2$,
where $v_c$ is the relative velocity of a collison. For small
bodies $v_c^2 \propto (e^2+\sin^2 i)/a$, where $a$, $e$, $i$ are
orbital elements. The value of $a$ is greater by a factor of 15
for the EKB than that for the MAB, a mean value of $(e^2+\sin^2 i)$ for
the EKB is less by a factor of 1.4 than that for the MAB. So, for
the EKB, $v_c^2$ is less by about a factor of $k \approx 20$, and
for the same composition of two colliding bodies the ratio $s_0$
needed for destruction of a larger body will be smaller by a
factor of $k$ than that for the MAB. Therefore, for the same composition of bodies, collisional lifetimes of TNOs are greater by about one or two orders of magnitude than those of asteroids. However, as it is more easy to
destruct icy EKB bodies than rocky bodies in the MAB, then the
above ratio $s_0$ may be much larger for the EKB, and the collisional
lifetimes of small bodies in the EKB may be of the same order that
those in the MAB. If some TNOs are porous, then it may be more difficult to destroy them than icy and rocky bodies. The total mass of SDOs 
moving in highly eccentric orbits between 40
and 200 AU is considered to be of the same order or greater than $M_{EKB}$. The mean energy of a collision of a SDO with an EKBO is greater (probably, on average, by a factor of 4) than that for two colliding EKBOs of the same
masses. Therefore, though SDOs spend a smaller part of their lifetimes at a distance $R<50$ AU, the probability of destruction of an EKBO (with
$30<a<50$ AU) by SDOs can be even more than that by EKBOs.

In contrast to the MAB, TNOs can get into the regions, from which they can migrate to orbits of planets, mainly due to gravitational interactions of with other TNOs, but not by collisions. The evolution of three gravitating bodies with masses equal to that of Pluto was investigated by Ipatov (1980) with the use of the spheres of action method (two two-bodies problems) and by Ipatov (1988) by numerical integration. The sphere of action (i.e., the Tisserand sphere) of radius $r_{s} \approx R (m/M_\odot)^{2/5}$ was used, where $R$ is a distance from the Sun, $M_\odot$ is the mass of the Sun, $m$ is the mass of an object. Basing on the results of numerical investigations of three and a hundred gravitating objects, Ipatov (1995a,b, 1998) made some appraisals of the evolution of a disk of TNOs. The used formulas for characteristic times elapsed up to close encounters are different from those for the \"Opik's approach used by other scientists. The case of variable inclinations was also considered. 
We obtained that the characteristic time elapsed up to a close encounter (up to $r_{s}$) of two test EKBOs with characteristic masses of the observed EKBOs ($m \sim 5 \cdot 10^{-12} M_\odot$) is equal to $7 \cdot 10^{10}$ yr. If there are about $7\cdot10^4$ EKBOs with $d \ge 100$ km, then during the last 4 Gyr a body with $a \approx 40$ AU takes part in about 3000 close (up to the Tisserand sphere) encounters with EKBOs with $d \ge 100$ km. The mean eccentricity of EKBOs could not reach its present value during the last 4 Gyr due to their mutual gravitational influence at the present mass of the EKB. Such growth could take place only for $M_{EKB}$ equal to several $m_\oplus$ (Ipatov, 1995a,b). The probability of that during the last 4 Gyr an EKBO with $d\ge100$ km collided with one of EKBOs of the same size equals to 0.005. If the number of EKBOs with $d\ge10$ km is greater than that with $d\ge100$ km by a factor of 100, then a probability of a collision during the last 4 Gyr of an EKBO with $d\ge100$ km with some EKBO with $d\ge10$ km equals to 0.15. As the mean eccentricity of EKBOs is about 0.1, variations in $a$ for colliding bodies can exceed several AU. Probably, less than 1\% of 100--km EKBOs had such large variations in $a$ due to collisions during the last 4 Gyr.

Our numerical estimates showed (Ipatov, 1995a) that the mean variation 
in $a$ at one close encounter is $\delta a \sim (1-3)\cdot10^{-5}a$ for $m=5
\cdot 10^{-12} M_\odot$ and mean eccentricities and inclinations of
the observed EKBOs. The variation in $a$ of an object located in the
middle of the EKB probably is proportional to $\delta a \sqrt{N_c}$
(where $N_c$ is the number of close encounters of this object with
large EKBOs) and usually does not exceed 0.1 AU during the last 4 Gyr
(all estimates are for the present mass of the EKB). For
many objects at the inner part of the belt (at $a < 39$ AU), $a$
could vary more monotonously and could decrease by more than 1 AU
during the last 4 Gyr. Such a decrease in $a$ may be a reason of that a region ($36 \le a \le 39$ AU) with small values of $e$ and $i$ is unpopulated, though, as it was shown by Duncan {\it et al.} (1995),
it is dynamically stable under the gravitational influence of
planets for the last 4 Gyr. At some very close encounters of EKBOs, variations in $a$ can be about several AU. Several percents of EKBOs could
take part in such very close encounters during the last 4 Gyr (Ipatov, 1998). 
The number of such close encounters is larger by an order of magnitude than
the corresponding number of collisions, and so the role of
gravitational interactions in variations of orbital elements is
larger than that of collisions. On average, during the last 4 Gyr a EKBO
had several close (up to $r_s$) encounters with Pluto and changed
its $a$ by $\sim 0.1$ AU at these encounters.

\centerline{}

\noindent
%\underline
{\bf Migration of Trans-Neptunian Objects under the Gravitational Influence of Planets}

%\centerline{}

Ipatov (1999), Ipatov and Henrard (2000) made computer runs of the evolution of orbits of some TNOs under the gravitational influence of planets.  The symplectic RMVS integrator worked out by Levison and Duncan (1994)
was used. The orbital evolution of one hundred test TNOs with various initial orbital
elements was considered. Initial values of semimajor axes
were varied from 35 to 50 AU. The considered time span usually
equaled 20 Myr. In some runs it reached 100--150 Myr. In contrast
to investigations by other authors, we were interested in the migration
of TNOs not only to the orbits of Neptune or Jupiter, but further
to the Earth's orbit. It was obtained in our runs that during
evolution four former TNOs reached Jupiter's orbit and for two
of them the perihelion distance decreased from 34 and 27.5 AU to
1.25 and 1.34 AU in 25 and 64 Myr, and these bodies were ejected
into hyperbolic orbits in 30 and 70 Myr, respectively. As orbital
elements were sampled with a time step equal to 20,000 yr, the
actual minimal values of a perihelion distance can be smaller than
those presented above.

Ipatov and Hahn (1999) investigated the evolution of orbits of
Jupiter-crossing objects close to the orbit of the object P/1996 R2
and obtained $p_{JE}\approx0.2$ and $\Delta t_E \approx 5000$ yr, where $p_{JE}$ is the portion of Jupiter-crossing objects that reached the Earth's orbit during their lifetimes, $\Delta t_E$ is the mean time during which a
Jupiter-crossing object crosses the Earth's orbit. At the initial
step equaled to 30 days, the value of $p_{JE}$ was smaller by a factor of several for the use of the RMVS integrator than that for the integrator by Bulirsh and Stoer 
%(1966). 
This difference was caused by the rare motion of bodies in the resonances, which decreased the perihelion distance, for the RMVS integrator. The use of the RMVS integrator could also cause an underestimate in the evaluation of probability for short-period comets to cross the Earth's orbit. Using this integrator, Levison and Duncan (1994) obtained about 0.5\% of their total number for $\Delta t_{E} = 10^3 - 10^5$ yr.

\centerline{}

\noindent
%\underline
{\bf The Number of Trans-Neptunian Objects Migrating to the Terrestrial Planets}

%\centerline{}

The number of TNOs migrating to the inner regions of the Solar System can be evaluated on the basis of simple formulas and the results of numerical integration. Let $N_J=P_N p_{JN}N$ be the number of former TNOs reaching Jupiter's orbit for the given time span $T$, where $N$ is the number of TNOs with $d \ge D$; $P_N$ is the fraction of TNOs leaving the belt and migrating to Neptune's orbit during $T$; and $p_{JN}$ is the fraction of Neptune-crossing objects which reach Jupiter's orbit for their lifetimes. Then the current number of Jupiter--crossers originated in the EKB equals $N_{Jn}=N_J \Delta t_J /T$, where $\Delta t_J$ is the average time during which the object crosses Jupiter's orbit. According to Duncan {\it et al.} ~(1995), the fraction $P_N$ of TNOs that left the EKB during $T$=4 Gyr under the influence of the giant planets is 0.1--0.2 and $p_{JN}$ = 0.34. As mutual gravitational influence of TNOs also takes place (Ipatov, 1998, 1999, 2000), we take this portion equal to 0.2. Hence, taking $T$=4 Gyr, $p_{JN}=0.34$, $\Delta t_J=0.2$ Myr, and $N = 10^{10}$ ($d \ge 1$ km), we have $N_J = 6.8 \cdot 10^8$ and $N_{Jn} = 3.4 \cdot 10^4$. Note that the number of Centaurs, coming from the EKB, turns out a bit less than $N_{Cn}=P_N T_C N /T$, where $T_C$ is the mean Centaurs' lifetime. For $N=7\cdot 10^4$ ($d \ge 100$ km), $P_N=0.2$, $T=4$ Gyr, and $T_C=10$ Myr, this yields $N_{Cn}=35$. The smaller $T_C$, the smaller $N_{Cn}$. For Uranus--crossers $T_C\sim10$ Myr, although for the Chiron-type objects $T_C\sim1$ Myr.

We can now estimate migration of TNOs to the Earth. Let the 
number of bodies approaching the Earth's orbit during the considered time span $T$ be $N_E = N_J p_{JE} = P_N p_{JN} p_{JE} N$, and about $N_{Jn} p_{JE} \Delta t_E / \Delta t_J$ former TNOs cross now both the orbits of Earth and Jupiter. Then the ratio of the number $N_{NE}$ of Earth-crossing objects (ECOs), which came from the EKB and are also Jupiter-crossers, to the total number $N_{ECO}$ of ECOs at the given time, is $P_{NE} = N_{NE} / N_{ECO} = P_N p_{JN} p_{JE} 
\Delta t_E N / (N_{ECO} T)$, where $N_{NE} = N_E \Delta t_E / T$. 
For $N=10^{10}$ ($d \ge 1$ km), $N_{ECO}=750$, $T=4$ Gyr,
$p_{JE}=0.2$, and $\Delta t_E \approx 5000$ yr, we have $P_{NE}
\approx 0.2$, $N_E \approx 1.4\cdot10^8$, $N_{NE} \approx 170$, i.e. about 20\% of 1-km Earth-crossers can be former TNOs, which now cross both the orbits of Jupiter and Earth. For larger values of $\Delta t_E$, the values of $P_{NE}$ are greater. $N_{NE}$ is smaller by a factor of 2 for $N=5\cdot 10^9$ (Jewitt and Fernandez, 2001) and it is much greater for $N=10^{11}$ (Jewitt, 1999) instead of $N=10^{10}$ as taken in the above estimate.
Some TNOs with semimajor axes $a>50$ AU can also migrate to the orbits of Jupiter and Earth. So the number of former TNOs which now cross both the orbits of the Earth and Jupiter is greater than $N_{NE}$.
Only a small part of Jupiter-crossing ECOs has been observed, because they use to move in distant orbits. Sometimes they move in 5:2, 7:3, 3:2 and other resonances with Jupiter (Ipatov and Hahn, 1999).
As it is more easy to destruct icy bodies than stone or metal
bodies (for example, on their way to the Earth, bodies can break up
at close encounters with planets and at collisions with small
bodies), the portion of TNOs among ECOs for bodies with $d\le 100$
m (for example, for Tunguska-size bodies) may be greater than that
for 1-km bodies, but small icy bodies disappear in the atmosphere
and cannot reach the surface of the Earth.

If one denotes the fraction of Jupiter-crossers reaching Earth-crossing orbits with aphelion distance $Q\le4.5$ AU as $P_Q$, and their mean residence time  in such orbits as $\Delta t_E^*$, then the number of such objects proves to be $N^*_{NE} = N_E P_Q \Delta t_E^* /T$. Valsecchi {\it et al.} (1995) obtained that the dynamical lifetime of comet P/Encke is about 0.35 Myr. The active lifetime of this comet is considered to be $\sim 10^3-10^4$ yr.
Wetherill (1988) estimated that extinct comets could provide 40\% of NEOs if one short-period comet evolved to an Encke-type orbit every $5\cdot10^4$ yr. Using the spheres of action method, Ipatov (1995b) investigated the orbital evolution of initially Jupiter-crossing bodies under the gravitational influence of all planets and found that $P_Q$ is about 0.04 (in these runs, several bodies got $Q<1$ AU). In the case of numerical integration, $P_Q$ is smaller. Harris and Bailey (1998) estimated $P_Q$ to be less than 0.0023. For about a hundred runs of evolution of Jupiter-crossing objects, Ipatov and Hahn (1999) didn't obtain Encke-type orbits. 
There is a possibility of decreasing $Q$ due to collisions with small bodies and destructions of comets during extremely close encounters with the terrestrial planets.
Weissman (1994) considered that gas sublimation from the cometary nucleus can also decrease $Q$. Asher {\it et al.} (2001) showed that the rate at which objects may be decoupled from Jupiter and attain orbits like NEOs is increased by a factor of four or five ($P_Q \sim 0.011$), producing $\sim 20\%$ of the observed NEO population, if nongraviational forces are included in integrations as impulsive effects.
For $N_{Jn}=30\,000$ (i.e., $d\ge1$ km), $\Delta t_E^*=0.35$ Myr, and $\Delta t_J=0.2$ Myr, we obtain $N_{En}^*\approx500$ for $P_Q=0.01$ and $N_{En}^*\approx100$ for $P_Q=0.002$. If a former Jupiter-crossing body moves in a typical 
Earth-crossing orbit, then $T_{En}$ and so $N_{En}$ can be even larger.
Therefore the number of former TNOs which now move inside Jupiter's orbit may be of the same order (or even more) as the number of former TNOs now crossing both the orbits of Jupiter and Earth. 

\centerline{}

\noindent
%\underline
{\bf Total Delivery of Celestial Bodies to the Earth}

%\centerline{}

Let us assess the potential collisions of celestial bodies with the terrestrial planets, taking the Earth as an example. 
The number of collisions of former TNOs, which now move in Jupiter-- and 
Earth--crossing orbits, with the Earth during time $T$ equals to $N_{col}^{TN} = N_E \Delta t_E/T_E$, where $T_E$ is a
characteristic time elapsed up to a collision of such ECO with the Earth. 
Note that the mean value of $T_E$ for the Jupiter-crossing ECOs is greater by a factor of several than that for other ECOs, so the portion of collisions of such objects is smaller than their portion among ECOs.
Bottke {\it et al.} (1994) obtained a mean collisional lifetime $T_{Ec}=134$ Myr for ECOs, and Ipatov (2000) using analytical 
formulas got $T_{Ec}=100$ Myr for the observed 417 ECOs. Ipatov's estimate $T_{Ec}=105$ Myr for 363 Apollo objects is in accord with $T_{Ec}=120$ Myr calculated by Dvorak and Pilat--Lohinger (1999) on the basis of the numerical integration of evolution of 54 Apollo objects. 
At $e=0.7$, $a=3.06$ AU ($P= 5.35$ yr), and $i$ varying from 0 to $30^\circ$, we have $T_E = 400$ Myr. Note that about half of observed Jupiter-family comets with $q<1$ AU have $P<5.35$ yr and for another half $P>5.35$ AU.

For $N=10^{10}$ ($d \ge 1$ km), $T$=4 Gyr, $p_{JE} = 0.2$, $\Delta t_E 
\approx 5000$ yr, and $T_E=400$ Myr, we obtain $N_{col}^{TN} \approx 
1750$, i.e., one collision per 2.3 Myr. For all ECOs with $d \ge 1$ km at $N_{ECO}=750$ and $T_{Ec}=100$ Myr, we have about 7.5 collisions per Myr. Therefore, the portion of former TNOs collided the Earth from Jupiter-crossing orbits among all 1-km bodies collided the Earth can be about several percents. The number of all 100-m ECOs is estimated to be about 70,000-160,000, so such objects collide the Earth on average ones in 600-1400 yr. For $d \ge 70$ m the frequency of collisions is greater by a factor of 2 than that for $d \ge 100$ m, and Tunguska-size objects collide the Earth once in several hundreds ($\sim 500$) years. 

Following the generally accepted concept that about half of NEOs are ECOs, we 
shall assess the late delivery of NEOs and former TNOs to the Earth. 
We assume that the number of NEOs with radii between 
$r$ and $r+dr$ equals $n(r) =c~dr/r^3$, where the dimension of $c$ is $r^2$. 
Integrating $n(r)\cdot(4/3)~\pi~\rho~r^3 = (4/3)~c~\pi~\rho~dr$ from $0$ to 
$r_{\max}$, we obtain the total mass of ECOs $M_{eco}=(4\pi/3)\rho~c~r_{\max}$, 
where $\rho$ is the mean density of  NEOs and $r_{\max}$ is the radius of the largest NEO ($\sim 20$ km) or the largest ECO ($\sim 5$ km). Below we consider $r_{\max}=20$ km. Integrating $n(r)= c~dr/r^3$ from some radius $r_*$ to $r_{\max}$, we obtain that the number $N_*$ of bodies with radius $r > r_*$ is $c~k_r/(2 r_*^2)$, where $k_r = 1- (r_*/r_{\max} )^2$. Obviously, for $r_*/r_{\max} = 0.1$, $k_r=0.99\approx1$. For example, if $r_*=0.5$ km, the mass $m_*= (4/3) \pi \rho r_*^3 =(\pi/6) k_\rho 10^{15} \approx 5.2\cdot10^{14} k_\rho$ g, where $k_\rho \sim 2$ is the ratio of the density $\rho$ of a NEO to that of water. Then the total mass of bodies
with $r>r_*$ turns out to be $M_{*}=(4\pi/3)\rho~c~(r_{\max}-r_*)$. 
On the other hand, if the mean mass of a body with $r>r_*$ equals $M_*/N_*$,  it will exceed $m_*$ by the factor of 
$(2 r_{\max} / r_*) r_{\max} / (r_{\max}+r_*) \approx 2 r_{\max} / r_*$. For example, for $\sim 1500$ NEOs with $r > 0.5$ km  it yields $M_{eco} 
\approx 750 (2r_{\max} / r_*) m_* \approx 6\cdot10^4 m_* \approx \pi 10^{19}k_\rho$ g.
If we suppose that $M_{eco}$ did not change significantly during $T=4$ Gyr, we obtain the total mass of bodies that collided with the Earth during $T$ to be equal to $M_{Et}=(T/T_{Ec})M_{eco} \approx 40 M_{eco}$. Assuming further that about $k_i=20$\% of them were icy bodies, the total mass of water delivered to the Earth during the last 4 Gyr turns out about $16\pi 10^{19} \approx 5\cdot10^{20}$ g, i.e., about $3.6\cdot10^{-4}$ of the mass of the Earth's hydrosphere. 

\centerline{}

\noindent{\bf
CONCLUSIONS}

%\centerline{}

Due to the gravitational influence of large trans-Neptunian objects and mutual collisions, some trans-Nep\-tu\-nian objects can get such orbits, from which they can be moved inside the Solar System under the gravitational influence of planets. The portion of former trans-Neptunian objects among Earth-crossing objects can be not small (up to 20\% or even more).

\centerline{}

\noindent
{\bf ACKNOWLEDGEMENTS}
\nolinebreak

%\centerline{}

This work was supported by the Belgian office for scientific, technical and cultural affairs (in 1999), the Russian Federal Program "Astronomy" (project 1.9.4.1), the Russian Foundation for Basic Research (project 01-02-17540), and INTAS (project 240 of INTAS Call 2000). 

\centerline{}

\noindent
{\bf REFERENCES}

%\centerline{}

%{
%\small

%}

\end{document}